\documentclass[%
 reprint,
 amsmath,amssymb,
 aps,
]{revtex4-1}

\usepackage{graphicx}
\usepackage{dcolumn}
\usepackage{bm}
\usepackage{keyval}

\usepackage{color}
\definecolor{deepblue}{rgb}{0,0,0.5}
\definecolor{deepred}{rgb}{0.6,0,0}
\definecolor{deepgreen}{rgb}{0,0.5,0}

\usepackage{listings}

\newcommand\pythonstyle{\lstset{
    language=Python,
    basicstyle=\footnotesize,
    otherkeywords={self},             
    keywordstyle=\footnotesize\color{deepblue},
    emph={__init__},          
    emphstyle=\footnotesize\color{deepred},    
    stringstyle=\color{deepgreen},
    frame=single,                         
    showstringspaces=false ,
    breaklines=true,
    numberstyle=\footnotesize,
    tabsize=3,
    breakatwhitespace=false
}}

\lstnewenvironment{python}[1]
{
\pythonstyle
\lstset{}
}
{}

\begin{document}

\preprint{APS/123-QED}

\title{Viscosity independent diffusion mediated by death and reproduction in biofilms}

\author{Arben Kalziqi}%
 
\affiliation{Georgia Tech
}%

\author{Siu Lung Ng}
\affiliation{%
 Georgia Tech
}%

\author{David Yanni}
\affiliation{%
 Georgia Tech
}%

\author{Gabi Steinbach}
\affiliation{%
 Georgia Tech
}%

\author{Brian K. Hammer}
\affiliation{%
 Georgia Tech
}%

\author{Peter J. Yunker}
\email{peter.yunker@physics.gatech.edu}
\affiliation{%
 Georgia Tech
}%

\date{\today}

\begin{abstract}
Bacterial biofilms, surface-attached communities of cells, are in some respects similar to colloidal solids; both are densely packed with non-zero yield stresses. However, unlike non-living materials, bacteria reproduce and die, breaking mechanical equilibrium and inducing collective dynamic responses. We report experiments and theory investigating the motion of immotile \textit{Vibrio cholerae}, which can kill each other and reproduce in biofilms. We vary viscosity by using bacterial variants that secrete different amounts of extracellular matrix polymers, but are otherwise identical. Unlike thermally-driven diffusion, in which diffusivity decreases with increased viscosity, we find that cellular motion mediated by death and reproduction is independent of viscosity over timescales relevant to bacterial reproduction. To understand this surprising result, we use two separate modeling approaches. First we perform explicitly mechanical simulations of one-dimensional chains of Voigt-Kelvin elements that can die and reproduce. Next, we perform an independent statistical approach, modeling Brownian motion with the classic Langevin equation under an effective temperature that depends on cellular division rate. The diffusion of cells in both approaches agrees quite well, supporting a kinetic interpretation for the effective temperature used here and developed in previous work. As the viscoelastic behavior of biofilms is believed to play a large role in their anomalous biological properties, such as antibiotic resistance, the independence of cellular diffusive motion --- important for biofilm growth and remodeling --- from viscoelastic properties likely holds ecological, medical, and industrial relevance.
\end{abstract}

\maketitle


Bacteria can grow in plankontic suspension or within biofilms \cite{RN12187}. These surface attached bacterial communities are similar to colloidal solids in many respects \cite{RN12193,RN11956}. Both are soft solids composed of densely-packed micron-sized objects. Biofilms even exhibit glassy phases, much like their colloidal counterparts \cite{RN12359,RN12155,henkes17}. However, unlike particles, bacteria reproduce and die; thus, even immotile bacteria that cannot swim are fundamentally active. Theory and experiments suggest that reproduction and death break mechanical equilibrium, driving fluctuations in cell motion and fluidizing biofilms \cite{RN12359,RN12155,RN12010,henkes17,malmi2018cell}. This active intra-biofilm diffusion occurs at high density, and within a highly viscoelastic material. The diffusion of cells via reproduction and death, therefore, is qualitatively unlike that experienced by thermal, non-living particles. 

Our understanding of thermal diffusion -- and specifically the Stokes-Einstein relation -- arises from a physical coincidence: the thermostatic collisions with solvent molecules driving Brownian motion are also responsible for viscous damping \cite{kubo1966fluctuation}. The fluctuation-dissipation relation then tells us that the force spectrum is determined by the friction spectrum; as a result, diffusivity is inversely proportional to viscosity. Conversely, in dense biofilms cellular diffusion is driven by local strain fields arising from either reproduction or death and lysis, while viscous damping is largely due to secreted extracellular matrix-forming polymers and cell-surface interactions \cite{RN12073,RN12360,RN12425,birjiniuk2014single,RN11957,RN13187}. Viscous and driving forces thus arise from completely separate sources in biofilms; how, then, is diffusion from cellular reproduction and death affected by viscosity?

Here, we measure the motion of cells in \textit{Vibrio cholerae} biofilms containing two variants that are genetically identical (``isogenic''), other than the genes responsible for encoding different Type VI Secretion System (``T6SS'') toxins. The T6SS is a contact-dependent toxin delivery system which allows these two otherwise-isogenic variants to kill each other when cells are in contact with one another. Genetically, we also vary whether the cells do or do not secrete extracellular matrix-forming polymers (``Matrix+'' and ``Matrix-'', respectively). We find that Matrix+ biofilms have a viscosity three times larger than that of Matrix- biofilms. Surprisingly, we find that desptie a large difference in viscosity, Matrix+ and Matrix- biofilms exhibit similar diffusivities.

We approach this problem with three complementary techniques. First, we track the motion of tracer beads embedded in biofilms with high and low viscosities via confocal microscopy. While tracking beads allows us to directly observe biofilm dynamics, beads only follow cellular motion in Matrix- biofilms; in Matrix+ biofilms bead dynamics deviate from cellular dynamics. To gain additional information about the average behavior of all cells, we use a previously validated method to extract this information from biofilm topography and independent mechanical measurements. Briefly, Risler, \textit{et al.}, developed a theory of homeostatic cellular films with reproduction and death, and predicted that an effective fluctuation-response relationship mediated by these opposing activities produces measurable effects on topography \cite{RN12359}. This prediction was experimentally validated in Kalziqi, \textit{et al.} \cite{RN13188}. Here, we combine topographic measurements with the generalized Stokes-Einstein relation to extract mean diffusivities. Finally, we model death and reproduction in biofilms with a series of Voigt-Kelvin dashpot-spring elements, and show through simulations and theory how viscosity impacts diffusion in such a system.

To explore the effect of the extracellular matrix on the motion and diffusion of cells within biofilms, we inoculated and incubated biofilms as described in \cite{RN13188}. To grow bioﬁlms containing isogenic mutual killers, we mixed (at a ratio of 1:1.4) two derivatives of a constitutive killer of C6706 \textit{V. cholerae} that are each genetically modified to use a different T6SS toxin from \textit{V. cholerae} strain 692-79  \cite{RN13188,RN12265,RN12520}. The mutual killing strains were also modified to express two different fluorescent proteins. Matrix- variants were genetically modified to produce no extracellular matrix \cite{RN12068}, whereas Matrix+ variants are wild type for extracellular matrix secretion (\textit{i.e.} they produce a ``natural'' amount of matrix product) \cite{RN12189}. All biofilms were grown by mixing two mutually-killing strains of \textit{V. cholerae} and placing a $1\mu$l inocula (roughly $8\times 10^5$ cells) onto a lysogeny broth agar plate at 37$^\circ$C.

First, we measured biofilm viscosity in the low-frequency limit via creep tests (see SI for more information) \cite{forgacs1998viscoelastic,boudarel2018towards,RN12014}. Matrix+ biofilms had an average viscosity of $43 \pm 5.1$ kPas (mean $ \pm $ standard deviation), just over three times higher than that of Matrix- biofilms ($14 \pm 6.2$ kPas) (Figure. 1a), demonstrating that the presence of extracellular matrix substantially modifies biofilm viscosity.

As a check on the impact of extracellular matrix on killing dynamics, we sought to measure the amount of killing that occurs in each type of biofilm. To do so, we first determined whether Matrix+/- biofilms have different levels of activity by examining the typical domain size of clonal patches. It was previously demonstrated that mutual killing bacteria undergo `Model A' coarsening in biofilms \cite{RN12189}; consequently, the size of clonal domains is related to the number of killing events. To measure the size of clonla domains, mutually-killing strains were incubated for 24 hours, and then visualized with fluorescence microscopy. As observed prior \cite{RN13188}, well-mixed, Matrix- mutual killers, which do not produce viscous extracellular matrix-forming polymers, separate to create a coarsened structure (Matrix- Fig. 1b and c).  Likewise, well-mixed Matrix+ mutual killers, which produce viscous extracellular matrix-forming polymers, exhibited similarly sized clonal domains (Matrix+, Fig, 1b and c). Thus, the amount of killing that occurred was not significantly altered by the presence or absence of matrix material.   

\begin{figure}
\centering
    \includegraphics[width=\linewidth]{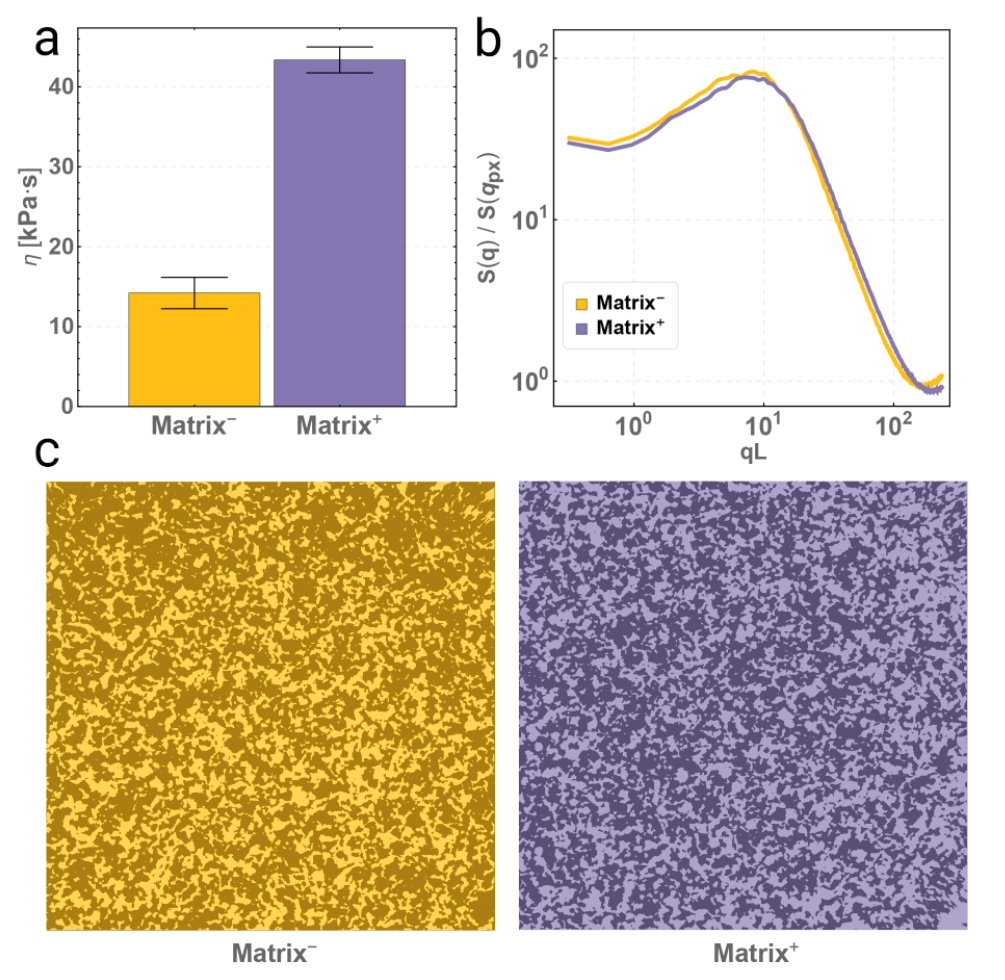}
    \caption{a. Directly measured viscosities for Matrix- and Matrix+ biofilms, shown with standard error.   The presence of extracellular matrix products increases the viscosity by roughly a factor of 3. b. Mean structure factors for Matrix- and Matrix+ biofilms show that the clonal size distribution is practically unaffected by the presence of the extracellular matrix, and thus the amount of killing is also unaffected.  c. Selected examples showing highly comparable strain-strain coarsening for Matrix- and Matrix+ biofilms.}
    \label{fig:Fig1}
\end{figure}

We next directly observed the motion of tracer beads in biofilms. 1-micron diameter polystyrene beads (FSFR004 Flash Red, Bangs Labs) were mixed with liquid culture immediately before the biofilm was inoculated. Beads were imaged every 3 minutes during 4.5 hours of growth with a Nikon A1R confocal microscope. It was previously shown that beads convect toward the top of growing biofilms at the agar-air interface \cite{RN13188}; thus, we measured the in-plane mean square displacement (MSD) (\textit{i.e.}, their mean square displacement parallel to the agar surface). In Matrix- biofilms, we found that tracer beads are highly mobile (Fig. 2a), exhibiting caged like dynamics on short lag times, and diffusive-like dynamics over longer lag times. We extract diffusion coefficients from the mean-squared displacement; Matrix- biofilms have a mean diffusivity of $0.63 \pm 0.17 \times 10^{-3} \mu m^{2}/s$. However, while they still exhibited diffusive dynamics, we found that beads did not effectively follow cellular dynamics in Matrix+ biofilms, likely due to their interactions with the extracellular matrix itself \cite{birjiniuk2014single} (see SI for additional information). Further, tracer beads are limited in the information they can provide about biofilm dynamics: tracer beads only probe their local neighborhoods, they convect away from the source of nutrients \cite{RN13188}, and it is unclear how the beads themselves may impact biofilm mechanics locally. Thus, while the motion of tracer beads strongly implies that immotile cells within a biofilm undergo significant, diffusive displacement due to reproduction and death, their inherent limitations motivate a new approach to measuring diffusivities that can capture the unperturbed dynamics of all cells.

\begin{figure}
\centering
    \includegraphics[width=\linewidth]{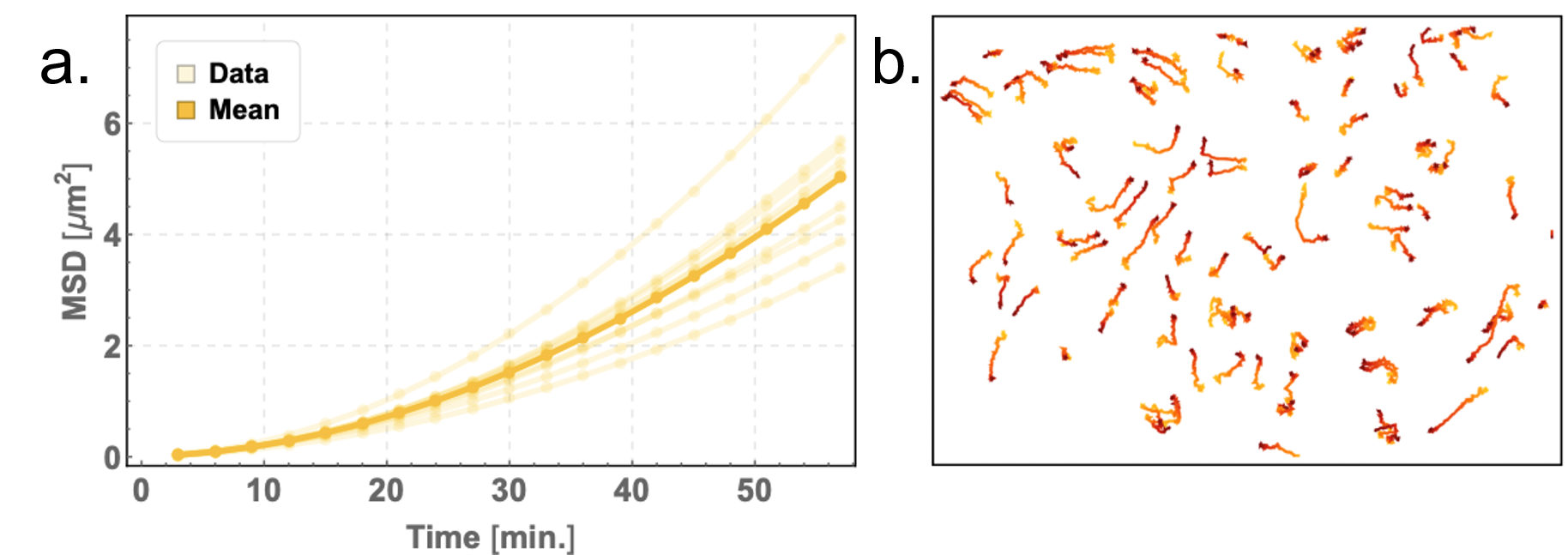}
    \caption{a. Measured lateral MSD for tracer beads inside a Matrix- biofilm. Light lines represent data from  individual measurements; the dark line represents the average MSD. b. Tracks of individual tracer beads colored by timepoint. Image is 100 microns wide and 74 microns tall.}
    \label{fig:Fig2}
\end{figure}

To do this, we turn to a previously-validated interferometry technique. In a recent paper \cite{RN13188}, we demonstrated that information about death and reproduction could be extracted from high resolution measurements of biofilm topography when coupled with mechanical measurements. That work built on previous theoretical results in which a Maxwell model of a tissue obeys an effective fluctuation-response relationship, permitting the calculation of integrated death and reproduction rates via an effective temperature ($T_{\textrm{eff}}$) in the low-frequency, long-wavelength limit \cite{RN12359}. This relationship was tested experimentally in \cite{RN13188} using Matrix- \textit{V. cholerae} biofilms; biofilms composed of mutual killing bacteria have $T_{\textrm{eff}}$ that are $\sim$18 times larger than those of biofilms composed of cells that cannot kill (p = $1.75 \times 10^{-4}$) \cite{RN13188}.

After 24 hours of incubation at 37$^\circ$C, the surface height profile was measured -- with O(nm) precision using a ZYGO NewView 8300 interferometer -- for 31 Matrix+ and 26 Matrix- biofilms. All measured biofilms are shaped like spherical caps with a central concave ellipsoidal dimple, often called the ``homeland'' \cite{hallatschek2007genetic}. Each biofilm topography is a superposition of such an ellipsoidal background and fluctuations due to cell-cell interactions. We fit and subtract an ellipsoidal background to each biofilm to obtain a fluctuation-topography in the vein of ref \cite{RN12359}, where the topography is determined by internal activity rather than initial and external conditions (Fig. 3 a). To relate topographies to effective temperatures (as in \cite{RN12359} and \cite{RN13188}), we must measure biofilm membrane tension and bending rigidity, which we did for Matrix+/- phenotypes (see SI).

\begin{figure}
\centering
    \includegraphics[width=\linewidth]{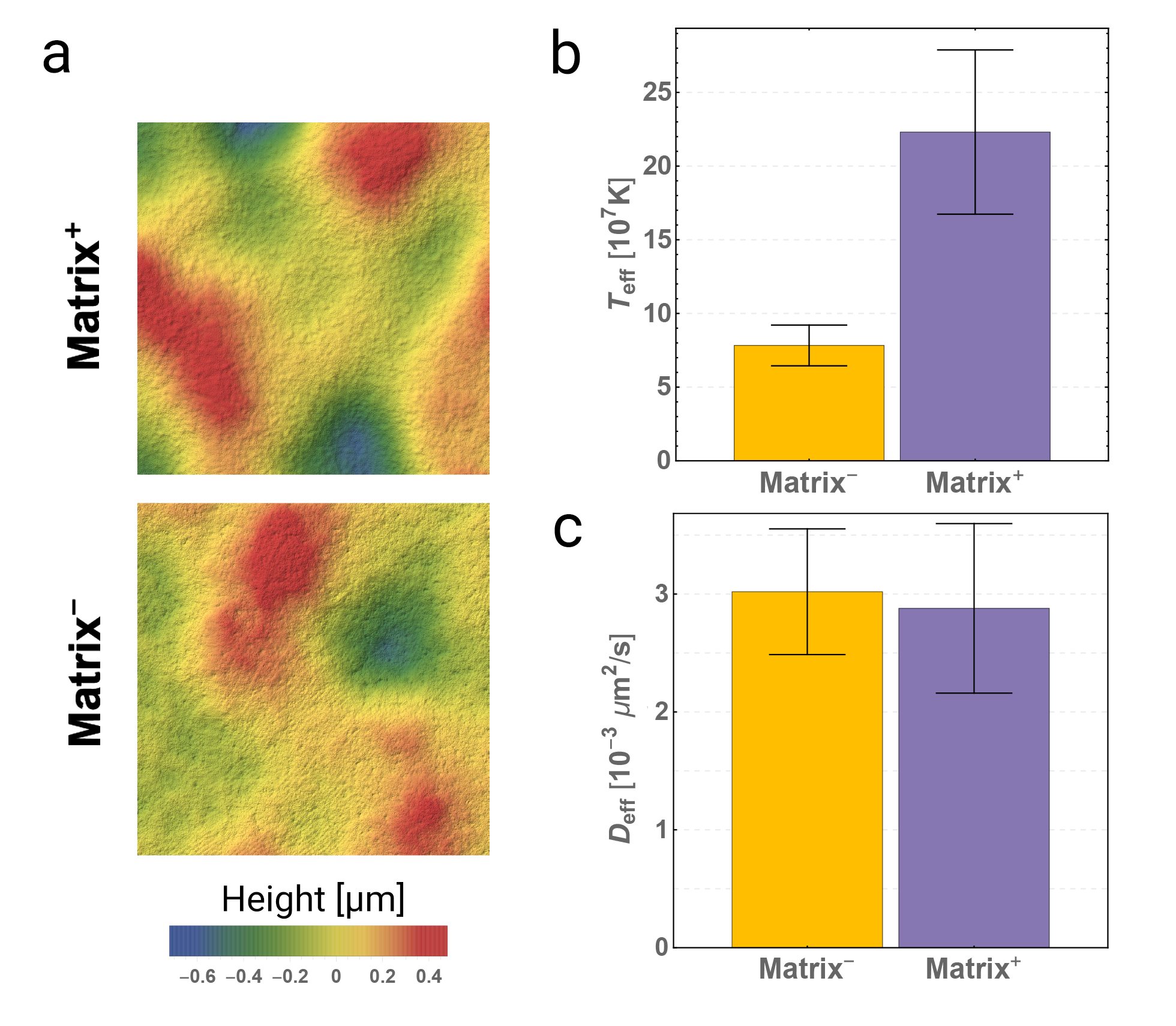}
    \caption{a. Demonstrative surface topographies of Matrix- and Matrix+ biofilm homelands measured via interferometry. b. While the topographies appear quite similar by eye, extracted effective temperatures (shown with standard error) are significantly different across many samples. c. However, diffusion constants calculated using the aforementioned effective temperatures and viscosities (shown with standard error) are nearly identical between Matrix- and Matrix+ samples.}
    \label{fig:Fig3}
\end{figure}

While Matrix+ and Matrix- biofilm topographies look superficially rough (Fig. 3a) \cite{C8BM00987B,PhysRevLett.121.238102}, these measurements reveal that Matrix+ biofilms have significantly larger effective temperatures than Matrix- biofilms ($p < 0.004$, Fig. 3b). We then convert measured effective temperatures into effective diffusivities, $D_{\textrm{eff}}$, using the generalized Stokes-Einstein relation in the low frequency limit (and independently-measured viscosities; see supplemental information for more details) (Fig. 3 c) \cite{mason1995tg}. While the effective temperature distributions are different depending on the presence of extracellular matrix, the distributions of effective diffusivities are very nearly identical ($p > 0.5$). The extracted mean is $D_{\textrm{eff}}=3.0 \pm 1.9 \times 10^{-3} \mu m^{2}/s$ and $D_{\textrm{eff}}=2.9 \pm 3.0 \times 10^{-3} \mu m^{2}/s$, for Matrix- and Matrix+ biofilms, respectively, each of which is reasonably close to the mean diffusivity extracted from Matrix- tracer bead experiments $D=0.63 \pm 0.17 \times 10^{-3} \mu m^{2}/s$. This agreement further supports the argument that biofilm topography directly relates to its underlying dynamics and mechanics.

\begin{figure*}[!tbh]
\centering
    \includegraphics[width=\textwidth]{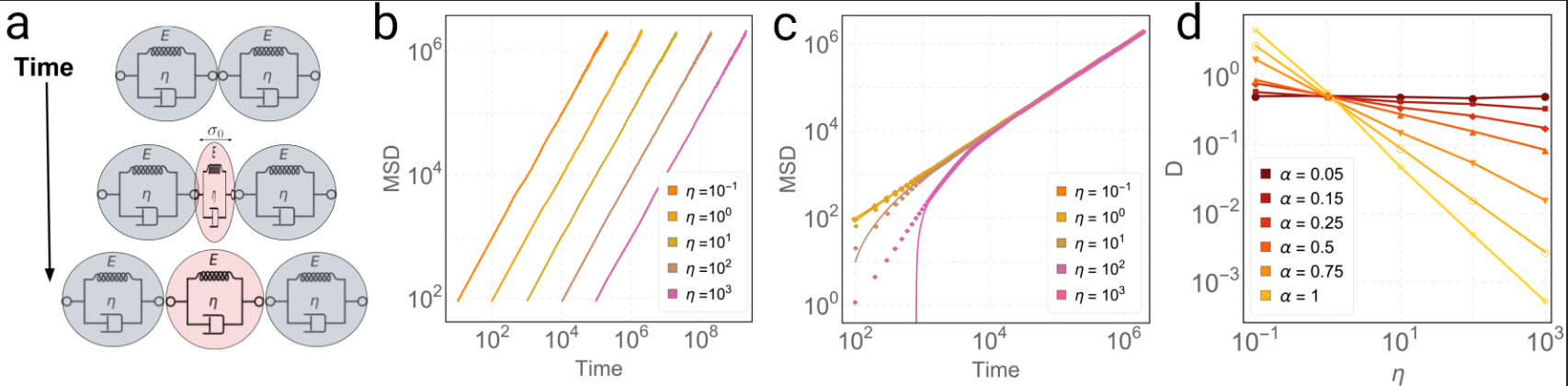}
    \caption{a. A visualization of the simulation setup. Cells are separated by Voigt-Kelvin elements and begin reproducing and lysing. Reproduction applies a stress $\sigma_0$ to neighboring cells. b. MSDs from simulations where $\alpha = 1$ (standard diffusion). c. MSDs from simulations where $\alpha = 0$ (free diffusion). The solid black line corresponds to an independent analytical prediction based on a generalized Langevin equation approach. d. Extracted diffusion constants as a function of viscosity from simulations with different values of $\alpha$.}
    \label{fig:Fig4}
\end{figure*}

To test this result, and gain insight into its universality, we perform event-driven, individual-based simulations of mutual killer cells in one-dimension. To capture the viscoelastic character of biofilms, we model them as chains of cells coupled by Voigt-Kelvin elements. Reproduction and death are assumed to be Poisson processes with activity rate $\lambda_{\textrm{act}}$ (Fig. 4a); the time-step between events is chosen from an exponential distribution $dt \sim e^{-\lambda_{\textrm{act}} t}$. Each event corresponds, with equal probability, to the step-strain resultant from reproduction or death of a cell at a random position in the biofilm, and as such imposes a step stress $\pm \sigma_0$ felt instantaneously throughout the biofilm. Finally, after each event the velocities and positions of all the cells are updated according to the current state of stress in the biofilm and the constitutive equations $\sigma(t) = E \epsilon(t)+\eta \frac{d\epsilon(t)}{dt} $, using backward-Euler integration, where $E$ is the elastic modulus, $\eta$ is viscosity and $\epsilon$ is strain.

The diffusion of cells in these simulations aligns well with the predictions of a simple modification to the classic Langevin equation model for the Brownian motion of a particle: $m \frac{dv}{dt}=-m\gamma v + R(t)$. Here, $\gamma$ is the inverse time scale associated with Stokes drag and $R(t)$ is a white noise term with zero mean and strength given by $\langle R(t)R(t+\tau)\rangle=2m\gamma k_B T_{\textrm{eff}}\delta(\tau)$, $k_B T_{\textrm{eff}} = U_0 \lambda_{\textrm{act}}/\gamma$, where $\lambda_{\textrm{act}}$ is the driving force activity rate, and $U_0$ is the energy scale of the driving force. When $\lambda_{\textrm{act}}$ is identified with the inverse time-scale associated with viscous damping ($\gamma$) and the energy scale is set to $U_0 = k_B T$, then the ordinary form of viscosity-dependent Brownian motion is recovered. On the other hand, when $\gamma$ is set to the viscoelastic relaxation rate ($E / \eta$),  and the energy scale is associated with cellular motion $U_0 = m (d \sigma_0 / \eta )^2$, where $d$ is the cellular diameter, we find that this Langevin equation agrees numerically with our simulations, and predicts diffusion to be independent of viscosity in the long time limit (Fig. 4c).

To further compare the simulations and theory, we set the simulated cellular reproduction and death rate to be a function of $E$ and $\eta$: $\lambda_{\textrm{act}} = \lambda_0 \left( \frac{E}{\eta} \right)^\alpha $ for a value of $\alpha$ between $0$ ($\lambda_{\textrm{act}}$ is totally independent of viscosity -- similar to actual bacterial reproduction rates) to $1$ ($\lambda_{\textrm{act}}$ has the traditional dependence on viscosity). Theory predicts that $D \propto -\alpha$; indeed, this is observed in simulations for several values of $\alpha$ (Fig. \ref{fig:Fig4}D). 

It is worth noting how surprising it is to recover the theoretically expected scaling between $D$ and $\alpha$ in simulations. The simulation is a viscoelastic mechanical model in which the only ingredients are death, reproduction and mechanical properties, and makes no explicit mention of effective temperatures or relaxation rates. Therein, we find that the mechanical properties play \textit{no} role in how cells move around in the long time limit, i.e. how the diffusion constant scales, \textit{unless} the rate of division and death is a function of those mechanical properties. And if the rate of division is set to be a function of mechanical properties, then the behavior of the diffusion constant in simulations is exactly as predicted by a theory that makes no explicit mention of viscoelasticity, mechanical properties, cell death, or cell reproduction. These represent two totally different approaches--- mechanical and statistical--- that yield the same result when translated into each others’ language. This observation offers more evidence that the effective temperature is not simply a thermodynamic analogue but actually relates to kinetics and the mechanical energy of particles, the same way ordinary temperature does \cite{RN12155,RN12359}. 
Thus, viscosity independent diffusion appears to be a natural consequence when the driving force and viscous damping arise from separate physical processes with different time scales.

Interestingly, the Langevin equation approach also predicts a diffusion constant of $D=\frac{1}{2} \lambda_{\textrm{act}} \ell^2$. Here $\lambda_{\textrm{act}}$ is the rate of lysis and reproduction, and $\ell$ is a characteristic length scale. A typical cell division rate for C6706 \textit{V. cholerae} is $\sim$20 minutes, so in the homeostatic limit the event rate can be roughly approximated as $\frac{1 \text{ event}}{10 \text{ min}}$. A typical cell length is about $1.0\, \mu \rm m$, so putting together a ``back of the envelope'' prediction we find $D \approx 1.5 \times 10^{-3} \mu \text{m}^2 \cdot s^{-1}$, on the same order of magnitude as effective diffusivities extracted from colony topography and from tracer beads in  Matrix- biofilms.


Through experiments, simulations, and theory, we show that diffusion mediated by death and reproduction is independent of viscosity. This surprising result arises from the separation of the origin of viscous relaxation and driving force time scales. Reproduction and death events induce step-strains that, in turn, induce stresses. These stresses relax much more rapidly than the time between step-strain events, making the material essentially a memory-less fluid on long time scales. The accumulated motion that a cell undergoes as a result of these strains amounts to a random walk governed entirely by the active driving force.

The results presented in this manuscript are likely of broad relevance. Biofilms that form in nature are typically polymicrobial, even featuring different taxa and species. Accordingly, bacteria have evolved many mechanisms for killing their competitors \cite{RN14451}, so this class of cellular diffusion is likely to be very common. Further, the mechanical properties of biofilms are critical for many other biological properties \cite{peterson2015viscoelasticity,RN11956,mah2001mechanisms,RN12195}; the independence of cellular diffusive motion -- important for biofilm growth and remodeling -- on biofilm mechanics suggests that biofilm viscoelasticity can vary without incurring a trade-off with regards to cellular diffusion.

\section*{Acknowledgements}
The authors would like to thank Joshua Weitz and Kurt Wiesenfeld for useful discussions. B.K.H. acknowledges funding from NSF (MCB-1149925). P.J.Y. acknowledges funding from Georgia Tech's Soft Matter Incubator. G.S. acknowledges funding from the German National Academy of Sciences Leopoldina (LPDS 2017-03).

\bibliography{refs}

\begin{thebibliography}{32}%
\makeatletter
\providecommand \@ifxundefined [1]{%
 \@ifx{#1\undefined}
}%
\providecommand \@ifnum [1]{%
 \ifnum #1\expandafter \@firstoftwo
 \else \expandafter \@secondoftwo
 \fi
}%
\providecommand \@ifx [1]{%
 \ifx #1\expandafter \@firstoftwo
 \else \expandafter \@secondoftwo
 \fi
}%
\providecommand \natexlab [1]{#1}%
\providecommand \enquote  [1]{``#1''}%
\providecommand \bibnamefont  [1]{#1}%
\providecommand \bibfnamefont [1]{#1}%
\providecommand \citenamefont [1]{#1}%
\providecommand \href@noop [0]{\@secondoftwo}%
\providecommand \href [0]{\begingroup \@sanitize@url \@href}%
\providecommand \@href[1]{\@@startlink{#1}\@@href}%
\providecommand \@@href[1]{\endgroup#1\@@endlink}%
\providecommand \@sanitize@url [0]{\catcode `\\12\catcode `\$12\catcode
  `\&12\catcode `\#12\catcode `\^12\catcode `\_12\catcode `\%12\relax}%
\providecommand \@@startlink[1]{}%
\providecommand \@@endlink[0]{}%
\providecommand \url  [0]{\begingroup\@sanitize@url \@url }%
\providecommand \@url [1]{\endgroup\@href {#1}{\urlprefix }}%
\providecommand \urlprefix  [0]{URL }%
\providecommand \Eprint [0]{\href }%
\providecommand \doibase [0]{http://dx.doi.org/}%
\providecommand \selectlanguage [0]{\@gobble}%
\providecommand \bibinfo  [0]{\@secondoftwo}%
\providecommand \bibfield  [0]{\@secondoftwo}%
\providecommand \translation [1]{[#1]}%
\providecommand \BibitemOpen [0]{}%
\providecommand \bibitemStop [0]{}%
\providecommand \bibitemNoStop [0]{.\EOS\space}%
\providecommand \EOS [0]{\spacefactor3000\relax}%
\providecommand \BibitemShut  [1]{\csname bibitem#1\endcsname}%
\let\auto@bib@innerbib\@empty
\bibitem [{\citenamefont {Hall-Stoodley}\ \emph {et~al.}(2004)\citenamefont
  {Hall-Stoodley}, \citenamefont {Costerton},\ and\ \citenamefont
  {Stoodley}}]{RN12187}%
  \BibitemOpen
  \bibfield  {author} {\bibinfo {author} {\bibfnamefont {L.}~\bibnamefont
  {Hall-Stoodley}}, \bibinfo {author} {\bibfnamefont {J.~W.}\ \bibnamefont
  {Costerton}}, \ and\ \bibinfo {author} {\bibfnamefont {P.}~\bibnamefont
  {Stoodley}},\ }\href {\doibase 10.1038/nrmicro821} {\bibfield  {journal}
  {\bibinfo  {journal} {Nat Rev Microbiol}\ }\textbf {\bibinfo {volume} {2}},\
  \bibinfo {pages} {95} (\bibinfo {year} {2004})}\BibitemShut {NoStop}%
\bibitem [{\citenamefont {Kovach}\ \emph {et~al.}(2017)\citenamefont {Kovach},
  \citenamefont {Davis-Fields}, \citenamefont {Irie}, \citenamefont {Jain},
  \citenamefont {Doorwar}, \citenamefont {Vuong}, \citenamefont {Dhamani},
  \citenamefont {Mohanty}, \citenamefont {Touhami},\ and\ \citenamefont
  {Gordon}}]{RN12193}%
  \BibitemOpen
  \bibfield  {author} {\bibinfo {author} {\bibfnamefont {K.}~\bibnamefont
  {Kovach}}, \bibinfo {author} {\bibfnamefont {M.}~\bibnamefont
  {Davis-Fields}}, \bibinfo {author} {\bibfnamefont {Y.}~\bibnamefont {Irie}},
  \bibinfo {author} {\bibfnamefont {K.}~\bibnamefont {Jain}}, \bibinfo {author}
  {\bibfnamefont {S.}~\bibnamefont {Doorwar}}, \bibinfo {author} {\bibfnamefont
  {K.}~\bibnamefont {Vuong}}, \bibinfo {author} {\bibfnamefont
  {N.}~\bibnamefont {Dhamani}}, \bibinfo {author} {\bibfnamefont
  {K.}~\bibnamefont {Mohanty}}, \bibinfo {author} {\bibfnamefont
  {A.}~\bibnamefont {Touhami}}, \ and\ \bibinfo {author} {\bibfnamefont
  {V.~D.}\ \bibnamefont {Gordon}},\ }\href {\doibase 10.1038/s41522-016-0007-9}
  {\bibfield  {journal} {\bibinfo  {journal} {NPJ Biofilms Microbiomes}\
  }\textbf {\bibinfo {volume} {3}},\ \bibinfo {pages} {1} (\bibinfo {year}
  {2017})}\BibitemShut {NoStop}%
\bibitem [{\citenamefont {Wilking}\ \emph {et~al.}(2011)\citenamefont
  {Wilking}, \citenamefont {Angelini}, \citenamefont {Seminara}, \citenamefont
  {Brenner},\ and\ \citenamefont {Weitz}}]{RN11956}%
  \BibitemOpen
  \bibfield  {author} {\bibinfo {author} {\bibfnamefont {J.~N.}\ \bibnamefont
  {Wilking}}, \bibinfo {author} {\bibfnamefont {T.~E.}\ \bibnamefont
  {Angelini}}, \bibinfo {author} {\bibfnamefont {A.}~\bibnamefont {Seminara}},
  \bibinfo {author} {\bibfnamefont {M.~P.}\ \bibnamefont {Brenner}}, \ and\
  \bibinfo {author} {\bibfnamefont {D.~A.}\ \bibnamefont {Weitz}},\ }\href
  {\doibase 10.1557/mrs.2011.71} {\bibfield  {journal} {\bibinfo  {journal}
  {Mrs Bulletin}\ }\textbf {\bibinfo {volume} {36}},\ \bibinfo {pages} {385}
  (\bibinfo {year} {2011})}\BibitemShut {NoStop}%
\bibitem [{\citenamefont {Risler}\ \emph {et~al.}(2015)\citenamefont {Risler},
  \citenamefont {Peilloux},\ and\ \citenamefont {Prost}}]{RN12359}%
  \BibitemOpen
  \bibfield  {author} {\bibinfo {author} {\bibfnamefont {T.}~\bibnamefont
  {Risler}}, \bibinfo {author} {\bibfnamefont {A.}~\bibnamefont {Peilloux}}, \
  and\ \bibinfo {author} {\bibfnamefont {J.}~\bibnamefont {Prost}},\ }\href
  {\doibase ARTN 258104 10.1103/PhysRevLett.115.258104} {\bibfield  {journal}
  {\bibinfo  {journal} {Physical Review Letters}\ }\textbf {\bibinfo {volume}
  {115}} (\bibinfo {year} {2015}),\ ARTN 258104
  10.1103/PhysRevLett.115.258104}\BibitemShut {NoStop}%
\bibitem [{\citenamefont {Ranft}\ \emph
  {et~al.}(2010{\natexlab{a}})\citenamefont {Ranft}, \citenamefont {Basan},
  \citenamefont {Elgeti}, \citenamefont {Joanny}, \citenamefont {Prost},\ and\
  \citenamefont {Julicher}}]{RN12155}%
  \BibitemOpen
  \bibfield  {author} {\bibinfo {author} {\bibfnamefont {J.}~\bibnamefont
  {Ranft}}, \bibinfo {author} {\bibfnamefont {M.}~\bibnamefont {Basan}},
  \bibinfo {author} {\bibfnamefont {J.}~\bibnamefont {Elgeti}}, \bibinfo
  {author} {\bibfnamefont {J.~F.}\ \bibnamefont {Joanny}}, \bibinfo {author}
  {\bibfnamefont {J.}~\bibnamefont {Prost}}, \ and\ \bibinfo {author}
  {\bibfnamefont {F.}~\bibnamefont {Julicher}},\ }\href {\doibase
  10.1073/pnas.1011086107} {\bibfield  {journal} {\bibinfo  {journal} {Proc
  Natl Acad Sci U S A}\ }\textbf {\bibinfo {volume} {107}},\ \bibinfo {pages}
  {20863} (\bibinfo {year} {2010}{\natexlab{a}})}\BibitemShut {NoStop}%
\bibitem [{\citenamefont {Matoz-Fernandez}\ \emph {et~al.}(2017)\citenamefont
  {Matoz-Fernandez}, \citenamefont {Martens}, \citenamefont {Sknepnek},
  \citenamefont {Barrat},\ and\ \citenamefont {Henkes}}]{henkes17}%
  \BibitemOpen
  \bibfield  {author} {\bibinfo {author} {\bibfnamefont {D.~A.}\ \bibnamefont
  {Matoz-Fernandez}}, \bibinfo {author} {\bibfnamefont {K.}~\bibnamefont
  {Martens}}, \bibinfo {author} {\bibfnamefont {R.}~\bibnamefont {Sknepnek}},
  \bibinfo {author} {\bibfnamefont {J.~L.}\ \bibnamefont {Barrat}}, \ and\
  \bibinfo {author} {\bibfnamefont {S.}~\bibnamefont {Henkes}},\ }\href
  {\doibase 10.1039/c6sm02580c} {\bibfield  {journal} {\bibinfo  {journal}
  {Soft Matter}\ }\textbf {\bibinfo {volume} {13}},\ \bibinfo {pages} {3205}
  (\bibinfo {year} {2017})}\BibitemShut {NoStop}%
\bibitem [{\citenamefont {Ranft}\ \emph
  {et~al.}(2010{\natexlab{b}})\citenamefont {Ranft}, \citenamefont {Basan},
  \citenamefont {Elgeti}, \citenamefont {Joanny}, \citenamefont {Prost},\ and\
  \citenamefont {Julicher}}]{RN12010}%
  \BibitemOpen
  \bibfield  {author} {\bibinfo {author} {\bibfnamefont {J.}~\bibnamefont
  {Ranft}}, \bibinfo {author} {\bibfnamefont {M.}~\bibnamefont {Basan}},
  \bibinfo {author} {\bibfnamefont {J.}~\bibnamefont {Elgeti}}, \bibinfo
  {author} {\bibfnamefont {J.~F.}\ \bibnamefont {Joanny}}, \bibinfo {author}
  {\bibfnamefont {J.}~\bibnamefont {Prost}}, \ and\ \bibinfo {author}
  {\bibfnamefont {F.}~\bibnamefont {Julicher}},\ }\href {\doibase
  10.1073/pnas.1011086107} {\bibfield  {journal} {\bibinfo  {journal}
  {Proceedings of the National Academy of Sciences of the United States of
  America}\ }\textbf {\bibinfo {volume} {107}},\ \bibinfo {pages} {20863}
  (\bibinfo {year} {2010}{\natexlab{b}})}\BibitemShut {NoStop}%
\bibitem [{\citenamefont {Malmi-Kakkada}\ \emph {et~al.}(2018)\citenamefont
  {Malmi-Kakkada}, \citenamefont {Li}, \citenamefont {Samanta}, \citenamefont
  {Sinha},\ and\ \citenamefont {Thirumalai}}]{malmi2018cell}%
  \BibitemOpen
  \bibfield  {author} {\bibinfo {author} {\bibfnamefont {A.~N.}\ \bibnamefont
  {Malmi-Kakkada}}, \bibinfo {author} {\bibfnamefont {X.}~\bibnamefont {Li}},
  \bibinfo {author} {\bibfnamefont {H.~S.}\ \bibnamefont {Samanta}}, \bibinfo
  {author} {\bibfnamefont {S.}~\bibnamefont {Sinha}}, \ and\ \bibinfo {author}
  {\bibfnamefont {D.}~\bibnamefont {Thirumalai}},\ }\href@noop {} {\bibfield
  {journal} {\bibinfo  {journal} {Physical Review X}\ }\textbf {\bibinfo
  {volume} {8}},\ \bibinfo {pages} {021025} (\bibinfo {year}
  {2018})}\BibitemShut {NoStop}%
\bibitem [{\citenamefont {Kubo}(1966)}]{kubo1966fluctuation}%
  \BibitemOpen
  \bibfield  {author} {\bibinfo {author} {\bibfnamefont {R.}~\bibnamefont
  {Kubo}},\ }in\ \href@noop {} {\emph {\bibinfo {booktitle} {Many-body
  theory}}}\ (\bibinfo {year} {1966})\ p.~\bibinfo {pages} {1}\BibitemShut
  {NoStop}%
\bibitem [{\citenamefont {Teschler}\ \emph {et~al.}(2015)\citenamefont
  {Teschler}, \citenamefont {Zamorano-Sanchez}, \citenamefont {Utada},
  \citenamefont {Warner}, \citenamefont {Wong}, \citenamefont {Linington},\
  and\ \citenamefont {Yildiz}}]{RN12073}%
  \BibitemOpen
  \bibfield  {author} {\bibinfo {author} {\bibfnamefont {J.~K.}\ \bibnamefont
  {Teschler}}, \bibinfo {author} {\bibfnamefont {D.}~\bibnamefont
  {Zamorano-Sanchez}}, \bibinfo {author} {\bibfnamefont {A.~S.}\ \bibnamefont
  {Utada}}, \bibinfo {author} {\bibfnamefont {C.~J.~A.}\ \bibnamefont
  {Warner}}, \bibinfo {author} {\bibfnamefont {G.~C.~L.}\ \bibnamefont {Wong}},
  \bibinfo {author} {\bibfnamefont {R.~G.}\ \bibnamefont {Linington}}, \ and\
  \bibinfo {author} {\bibfnamefont {F.~H.}\ \bibnamefont {Yildiz}},\ }\href
  {\doibase 10.1038/nrmicro3433} {\bibfield  {journal} {\bibinfo  {journal}
  {Nature Reviews Microbiology}\ }\textbf {\bibinfo {volume} {13}},\ \bibinfo
  {pages} {255} (\bibinfo {year} {2015})}\BibitemShut {NoStop}%
\bibitem [{\citenamefont {Yan}\ \emph {et~al.}(2017)\citenamefont {Yan},
  \citenamefont {Nadell}, \citenamefont {Stone}, \citenamefont {Wingreen},\
  and\ \citenamefont {Bassler}}]{RN12360}%
  \BibitemOpen
  \bibfield  {author} {\bibinfo {author} {\bibfnamefont {J.}~\bibnamefont
  {Yan}}, \bibinfo {author} {\bibfnamefont {C.~D.}\ \bibnamefont {Nadell}},
  \bibinfo {author} {\bibfnamefont {H.~A.}\ \bibnamefont {Stone}}, \bibinfo
  {author} {\bibfnamefont {N.~S.}\ \bibnamefont {Wingreen}}, \ and\ \bibinfo
  {author} {\bibfnamefont {B.~L.}\ \bibnamefont {Bassler}},\ }\href {\doibase
  10.1038/s41467-017-00401-1} {\bibfield  {journal} {\bibinfo  {journal} {Nat
  Commun}\ }\textbf {\bibinfo {volume} {8}},\ \bibinfo {pages} {327} (\bibinfo
  {year} {2017})}\BibitemShut {NoStop}%
\bibitem [{\citenamefont {Rodesney}\ \emph
  {et~al.}(2017{\natexlab{a}})\citenamefont {Rodesney}, \citenamefont {Roman},
  \citenamefont {Dhamani}, \citenamefont {Cooley}, \citenamefont {Touhami},\
  and\ \citenamefont {Gordon}}]{RN12425}%
  \BibitemOpen
  \bibfield  {author} {\bibinfo {author} {\bibfnamefont {C.~A.}\ \bibnamefont
  {Rodesney}}, \bibinfo {author} {\bibfnamefont {B.}~\bibnamefont {Roman}},
  \bibinfo {author} {\bibfnamefont {N.}~\bibnamefont {Dhamani}}, \bibinfo
  {author} {\bibfnamefont {B.~J.}\ \bibnamefont {Cooley}}, \bibinfo {author}
  {\bibfnamefont {A.}~\bibnamefont {Touhami}}, \ and\ \bibinfo {author}
  {\bibfnamefont {V.~D.}\ \bibnamefont {Gordon}},\ }\href {\doibase
  10.1073/pnas.1703255114} {\bibfield  {journal} {\bibinfo  {journal} {Proc
  Natl Acad Sci U S A}\ }\textbf {\bibinfo {volume} {114}},\ \bibinfo {pages}
  {5906} (\bibinfo {year} {2017}{\natexlab{a}})}\BibitemShut {NoStop}%
\bibitem [{\citenamefont {Birjiniuk}\ \emph {et~al.}(2014)\citenamefont
  {Birjiniuk}, \citenamefont {Billings}, \citenamefont {Nance}, \citenamefont
  {Hanes}, \citenamefont {Ribbeck},\ and\ \citenamefont
  {Doyle}}]{birjiniuk2014single}%
  \BibitemOpen
  \bibfield  {author} {\bibinfo {author} {\bibfnamefont {A.}~\bibnamefont
  {Birjiniuk}}, \bibinfo {author} {\bibfnamefont {N.}~\bibnamefont {Billings}},
  \bibinfo {author} {\bibfnamefont {E.}~\bibnamefont {Nance}}, \bibinfo
  {author} {\bibfnamefont {J.}~\bibnamefont {Hanes}}, \bibinfo {author}
  {\bibfnamefont {K.}~\bibnamefont {Ribbeck}}, \ and\ \bibinfo {author}
  {\bibfnamefont {P.~S.}\ \bibnamefont {Doyle}},\ }\href@noop {} {\bibfield
  {journal} {\bibinfo  {journal} {New Journal of Physics}\ }\textbf {\bibinfo
  {volume} {16}},\ \bibinfo {pages} {085014} (\bibinfo {year}
  {2014})}\BibitemShut {NoStop}%
\bibitem [{\citenamefont {Seminara}\ \emph {et~al.}(2012)\citenamefont
  {Seminara}, \citenamefont {Angelini}, \citenamefont {Wilking}, \citenamefont
  {Vlamakis}, \citenamefont {Ebrahim}, \citenamefont {Kolter}, \citenamefont
  {Weitz},\ and\ \citenamefont {Brenner}}]{RN11957}%
  \BibitemOpen
  \bibfield  {author} {\bibinfo {author} {\bibfnamefont {A.}~\bibnamefont
  {Seminara}}, \bibinfo {author} {\bibfnamefont {T.~E.}\ \bibnamefont
  {Angelini}}, \bibinfo {author} {\bibfnamefont {J.~N.}\ \bibnamefont
  {Wilking}}, \bibinfo {author} {\bibfnamefont {H.}~\bibnamefont {Vlamakis}},
  \bibinfo {author} {\bibfnamefont {S.}~\bibnamefont {Ebrahim}}, \bibinfo
  {author} {\bibfnamefont {R.}~\bibnamefont {Kolter}}, \bibinfo {author}
  {\bibfnamefont {D.~A.}\ \bibnamefont {Weitz}}, \ and\ \bibinfo {author}
  {\bibfnamefont {M.~P.}\ \bibnamefont {Brenner}},\ }\href {\doibase
  10.1073/pnas.1109261108} {\bibfield  {journal} {\bibinfo  {journal}
  {Proceedings of the National Academy of Sciences of the United States of
  America}\ }\textbf {\bibinfo {volume} {109}},\ \bibinfo {pages} {1116}
  (\bibinfo {year} {2012})}\BibitemShut {NoStop}%
\bibitem [{\citenamefont {Pentz}\ \emph {et~al.}(2018)\citenamefont {Pentz},
  \citenamefont {Márquez-Zacarías}, \citenamefont {Yunker}, \citenamefont
  {Libby},\ and\ \citenamefont {Ratcliff}}]{RN13187}%
  \BibitemOpen
  \bibfield  {author} {\bibinfo {author} {\bibfnamefont {J.~T.}\ \bibnamefont
  {Pentz}}, \bibinfo {author} {\bibfnamefont {P.}~\bibnamefont
  {Márquez-Zacarías}}, \bibinfo {author} {\bibfnamefont {P.~J.}\ \bibnamefont
  {Yunker}}, \bibinfo {author} {\bibfnamefont {E.}~\bibnamefont {Libby}}, \
  and\ \bibinfo {author} {\bibfnamefont {W.~C.}\ \bibnamefont {Ratcliff}},\
  }\href {\doibase 10.1101/255307} {\bibfield  {journal} {\bibinfo  {journal}
  {bioRxiv}\ } (\bibinfo {year} {2018}),\ 10.1101/255307}\BibitemShut {NoStop}%
\bibitem [{\citenamefont {Kalziqi}\ \emph {et~al.}(2018)\citenamefont
  {Kalziqi}, \citenamefont {Yanni}, \citenamefont {Thomas}, \citenamefont {Ng},
  \citenamefont {Vivek}, \citenamefont {Hammer},\ and\ \citenamefont
  {Yunker}}]{RN13188}%
  \BibitemOpen
  \bibfield  {author} {\bibinfo {author} {\bibfnamefont {A.}~\bibnamefont
  {Kalziqi}}, \bibinfo {author} {\bibfnamefont {D.}~\bibnamefont {Yanni}},
  \bibinfo {author} {\bibfnamefont {J.}~\bibnamefont {Thomas}}, \bibinfo
  {author} {\bibfnamefont {S.~L.}\ \bibnamefont {Ng}}, \bibinfo {author}
  {\bibfnamefont {S.}~\bibnamefont {Vivek}}, \bibinfo {author} {\bibfnamefont
  {B.~K.}\ \bibnamefont {Hammer}}, \ and\ \bibinfo {author} {\bibfnamefont
  {P.~J.}\ \bibnamefont {Yunker}},\ }\href {\doibase
  10.1103/PhysRevLett.120.018101} {\bibfield  {journal} {\bibinfo  {journal}
  {Phys Rev Lett}\ }\textbf {\bibinfo {volume} {120}},\ \bibinfo {pages}
  {018101} (\bibinfo {year} {2018})}\BibitemShut {NoStop}%
\bibitem [{\citenamefont {Thomas}\ \emph {et~al.}(2017)\citenamefont {Thomas},
  \citenamefont {Watve}, \citenamefont {Ratcliff},\ and\ \citenamefont
  {Hammer}}]{RN12265}%
  \BibitemOpen
  \bibfield  {author} {\bibinfo {author} {\bibfnamefont {J.}~\bibnamefont
  {Thomas}}, \bibinfo {author} {\bibfnamefont {S.~S.}\ \bibnamefont {Watve}},
  \bibinfo {author} {\bibfnamefont {W.~C.}\ \bibnamefont {Ratcliff}}, \ and\
  \bibinfo {author} {\bibfnamefont {B.~K.}\ \bibnamefont {Hammer}},\ }\href
  {\doibase 10.1128/mBio.00654-17} {\bibfield  {journal} {\bibinfo  {journal}
  {MBio}\ }\textbf {\bibinfo {volume} {8}} (\bibinfo {year} {2017}),\
  10.1128/mBio.00654-17}\BibitemShut {NoStop}%
\bibitem [{\citenamefont {Watve}\ \emph {et~al.}(2015)\citenamefont {Watve},
  \citenamefont {Thomas},\ and\ \citenamefont {Hammer}}]{RN12520}%
  \BibitemOpen
  \bibfield  {author} {\bibinfo {author} {\bibfnamefont {S.~S.}\ \bibnamefont
  {Watve}}, \bibinfo {author} {\bibfnamefont {J.}~\bibnamefont {Thomas}}, \
  and\ \bibinfo {author} {\bibfnamefont {B.~K.}\ \bibnamefont {Hammer}},\
  }\href {\doibase 10.1371/journal.pone.0138834} {\bibfield  {journal}
  {\bibinfo  {journal} {PLoS One}\ }\textbf {\bibinfo {volume} {10}},\ \bibinfo
  {pages} {e0138834} (\bibinfo {year} {2015})}\BibitemShut {NoStop}%
\bibitem [{\citenamefont {Hammer}\ and\ \citenamefont
  {Bassler}(2004)}]{RN12068}%
  \BibitemOpen
  \bibfield  {author} {\bibinfo {author} {\bibfnamefont {B.~K.}\ \bibnamefont
  {Hammer}}\ and\ \bibinfo {author} {\bibfnamefont {B.~L.}\ \bibnamefont
  {Bassler}},\ }\href {\doibase 10.1046/j.1365-2958.2004.03939.x} {\bibfield
  {journal} {\bibinfo  {journal} {Molecular Microbiology}\ }\textbf {\bibinfo
  {volume} {51}},\ \bibinfo {pages} {1521} (\bibinfo {year}
  {2004})}\BibitemShut {NoStop}%
\bibitem [{\citenamefont {McNally}\ \emph {et~al.}(2017)\citenamefont
  {McNally}, \citenamefont {Bernardy}, \citenamefont {Thomas}, \citenamefont
  {Kalziqi}, \citenamefont {Pentz}, \citenamefont {Brown}, \citenamefont
  {Hammer}, \citenamefont {Yunker},\ and\ \citenamefont {Ratcliff}}]{RN12189}%
  \BibitemOpen
  \bibfield  {author} {\bibinfo {author} {\bibfnamefont {L.}~\bibnamefont
  {McNally}}, \bibinfo {author} {\bibfnamefont {E.}~\bibnamefont {Bernardy}},
  \bibinfo {author} {\bibfnamefont {J.}~\bibnamefont {Thomas}}, \bibinfo
  {author} {\bibfnamefont {A.}~\bibnamefont {Kalziqi}}, \bibinfo {author}
  {\bibfnamefont {J.}~\bibnamefont {Pentz}}, \bibinfo {author} {\bibfnamefont
  {S.~P.}\ \bibnamefont {Brown}}, \bibinfo {author} {\bibfnamefont {B.~K.}\
  \bibnamefont {Hammer}}, \bibinfo {author} {\bibfnamefont {P.~J.}\
  \bibnamefont {Yunker}}, \ and\ \bibinfo {author} {\bibfnamefont {W.~C.}\
  \bibnamefont {Ratcliff}},\ }\href {\doibase 10.1038/ncomms14371} {\bibfield
  {journal} {\bibinfo  {journal} {Nat Commun}\ }\textbf {\bibinfo {volume}
  {8}},\ \bibinfo {pages} {14371} (\bibinfo {year} {2017})}\BibitemShut
  {NoStop}%
\bibitem [{\citenamefont {Forgacs}\ \emph {et~al.}(1998)\citenamefont
  {Forgacs}, \citenamefont {Foty}, \citenamefont {Shafrir},\ and\ \citenamefont
  {Steinberg}}]{forgacs1998viscoelastic}%
  \BibitemOpen
  \bibfield  {author} {\bibinfo {author} {\bibfnamefont {G.}~\bibnamefont
  {Forgacs}}, \bibinfo {author} {\bibfnamefont {R.~A.}\ \bibnamefont {Foty}},
  \bibinfo {author} {\bibfnamefont {Y.}~\bibnamefont {Shafrir}}, \ and\
  \bibinfo {author} {\bibfnamefont {M.~S.}\ \bibnamefont {Steinberg}},\
  }\href@noop {} {\bibfield  {journal} {\bibinfo  {journal} {Biophysical
  journal}\ }\textbf {\bibinfo {volume} {74}},\ \bibinfo {pages} {2227}
  (\bibinfo {year} {1998})}\BibitemShut {NoStop}%
\bibitem [{\citenamefont {Boudarel}\ \emph {et~al.}(2018)\citenamefont
  {Boudarel}, \citenamefont {Mathias}, \citenamefont {Blaysat},\ and\
  \citenamefont {Gr{\'e}diac}}]{boudarel2018towards}%
  \BibitemOpen
  \bibfield  {author} {\bibinfo {author} {\bibfnamefont {H.}~\bibnamefont
  {Boudarel}}, \bibinfo {author} {\bibfnamefont {J.-D.}\ \bibnamefont
  {Mathias}}, \bibinfo {author} {\bibfnamefont {B.}~\bibnamefont {Blaysat}}, \
  and\ \bibinfo {author} {\bibfnamefont {M.}~\bibnamefont {Gr{\'e}diac}},\
  }\href@noop {} {\bibfield  {journal} {\bibinfo  {journal} {NPJ biofilms and
  microbiomes}\ }\textbf {\bibinfo {volume} {4}},\ \bibinfo {pages} {17}
  (\bibinfo {year} {2018})}\BibitemShut {NoStop}%
\bibitem [{\citenamefont {Vigers}\ and\ \citenamefont
  {Wilking}(2016)}]{RN12014}%
  \BibitemOpen
  \bibfield  {author} {\bibinfo {author} {\bibfnamefont {M.~P.}\ \bibnamefont
  {Vigers}}\ and\ \bibinfo {author} {\bibfnamefont {J.~N.}\ \bibnamefont
  {Wilking}},\ }\href {<Go to ISI>://WOS:000375093800347} {\bibfield  {journal}
  {\bibinfo  {journal} {Biophysical Journal}\ }\textbf {\bibinfo {volume}
  {110}},\ \bibinfo {pages} {172a} (\bibinfo {year} {2016})}\BibitemShut
  {NoStop}%
\bibitem [{\citenamefont {Hallatschek}\ \emph {et~al.}(2007)\citenamefont
  {Hallatschek}, \citenamefont {Hersen}, \citenamefont {Ramanathan},\ and\
  \citenamefont {Nelson}}]{hallatschek2007genetic}%
  \BibitemOpen
  \bibfield  {author} {\bibinfo {author} {\bibfnamefont {O.}~\bibnamefont
  {Hallatschek}}, \bibinfo {author} {\bibfnamefont {P.}~\bibnamefont {Hersen}},
  \bibinfo {author} {\bibfnamefont {S.}~\bibnamefont {Ramanathan}}, \ and\
  \bibinfo {author} {\bibfnamefont {D.~R.}\ \bibnamefont {Nelson}},\
  }\href@noop {} {\bibfield  {journal} {\bibinfo  {journal} {Proceedings of the
  National Academy of Sciences}\ }\textbf {\bibinfo {volume} {104}},\ \bibinfo
  {pages} {19926} (\bibinfo {year} {2007})}\BibitemShut {NoStop}%
\bibitem [{\citenamefont {Falcón~García}\ \emph {et~al.}(2019)\citenamefont
  {Falcón~García}, \citenamefont {Stangl}, \citenamefont {Götz},
  \citenamefont {Zhao}, \citenamefont {Sieber}, \citenamefont {Opitz},\ and\
  \citenamefont {Lieleg}}]{C8BM00987B}%
  \BibitemOpen
  \bibfield  {author} {\bibinfo {author} {\bibfnamefont {C.}~\bibnamefont
  {Falcón~García}}, \bibinfo {author} {\bibfnamefont {F.}~\bibnamefont
  {Stangl}}, \bibinfo {author} {\bibfnamefont {A.}~\bibnamefont {Götz}},
  \bibinfo {author} {\bibfnamefont {W.}~\bibnamefont {Zhao}}, \bibinfo {author}
  {\bibfnamefont {S.~A.}\ \bibnamefont {Sieber}}, \bibinfo {author}
  {\bibfnamefont {M.}~\bibnamefont {Opitz}}, \ and\ \bibinfo {author}
  {\bibfnamefont {O.}~\bibnamefont {Lieleg}},\ }\href {\doibase
  10.1039/C8BM00987B} {\bibfield  {journal} {\bibinfo  {journal} {Biomater.
  Sci.}\ ,\ } (\bibinfo {year} {2019})}\BibitemShut {NoStop}%
\bibitem [{\citenamefont {Williamson}\ and\ \citenamefont
  {Salbreux}(2018)}]{PhysRevLett.121.238102}%
  \BibitemOpen
  \bibfield  {author} {\bibinfo {author} {\bibfnamefont {J.~J.}\ \bibnamefont
  {Williamson}}\ and\ \bibinfo {author} {\bibfnamefont {G.}~\bibnamefont
  {Salbreux}},\ }\href {\doibase 10.1103/PhysRevLett.121.238102} {\bibfield
  {journal} {\bibinfo  {journal} {Phys. Rev. Lett.}\ }\textbf {\bibinfo
  {volume} {121}},\ \bibinfo {pages} {238102} (\bibinfo {year}
  {2018})}\BibitemShut {NoStop}%
\bibitem [{\citenamefont {Mason}(1995)}]{mason1995tg}%
  \BibitemOpen
  \bibfield  {author} {\bibinfo {author} {\bibfnamefont {T.}~\bibnamefont
  {Mason}},\ }\href@noop {} {\bibfield  {journal} {\bibinfo  {journal} {Phys.
  Rev. Lett.}\ }\textbf {\bibinfo {volume} {74}},\ \bibinfo {pages} {1250}
  (\bibinfo {year} {1995})}\BibitemShut {NoStop}%
\bibitem [{\citenamefont {Garcia-Bayona}\ and\ \citenamefont
  {Comstock}(2018)}]{RN14451}%
  \BibitemOpen
  \bibfield  {author} {\bibinfo {author} {\bibfnamefont {L.}~\bibnamefont
  {Garcia-Bayona}}\ and\ \bibinfo {author} {\bibfnamefont {L.~E.}\ \bibnamefont
  {Comstock}},\ }\href {\doibase 10.1126/science.aat2456} {\bibfield  {journal}
  {\bibinfo  {journal} {Science}\ }\textbf {\bibinfo {volume} {361}} (\bibinfo
  {year} {2018}),\ 10.1126/science.aat2456}\BibitemShut {NoStop}%
\bibitem [{\citenamefont {Peterson}\ \emph {et~al.}(2015)\citenamefont
  {Peterson}, \citenamefont {He}, \citenamefont {Ren}, \citenamefont {Zerdoum},
  \citenamefont {Libera}, \citenamefont {Sharma}, \citenamefont
  {Van~Winkelhoff}, \citenamefont {Neut}, \citenamefont {Stoodley},
  \citenamefont {Van Der~Mei} \emph {et~al.}}]{peterson2015viscoelasticity}%
  \BibitemOpen
  \bibfield  {author} {\bibinfo {author} {\bibfnamefont {B.~W.}\ \bibnamefont
  {Peterson}}, \bibinfo {author} {\bibfnamefont {Y.}~\bibnamefont {He}},
  \bibinfo {author} {\bibfnamefont {Y.}~\bibnamefont {Ren}}, \bibinfo {author}
  {\bibfnamefont {A.}~\bibnamefont {Zerdoum}}, \bibinfo {author} {\bibfnamefont
  {M.~R.}\ \bibnamefont {Libera}}, \bibinfo {author} {\bibfnamefont {P.~K.}\
  \bibnamefont {Sharma}}, \bibinfo {author} {\bibfnamefont {A.-J.}\
  \bibnamefont {Van~Winkelhoff}}, \bibinfo {author} {\bibfnamefont
  {D.}~\bibnamefont {Neut}}, \bibinfo {author} {\bibfnamefont {P.}~\bibnamefont
  {Stoodley}}, \bibinfo {author} {\bibfnamefont {H.~C.}\ \bibnamefont {Van
  Der~Mei}},  \emph {et~al.},\ }\href@noop {} {\bibfield  {journal} {\bibinfo
  {journal} {FEMS microbiology reviews}\ }\textbf {\bibinfo {volume} {39}},\
  \bibinfo {pages} {234} (\bibinfo {year} {2015})}\BibitemShut {NoStop}%
\bibitem [{\citenamefont {Mah}\ and\ \citenamefont
  {O'toole}(2001)}]{mah2001mechanisms}%
  \BibitemOpen
  \bibfield  {author} {\bibinfo {author} {\bibfnamefont {T.-F.~C.}\
  \bibnamefont {Mah}}\ and\ \bibinfo {author} {\bibfnamefont {G.~A.}\
  \bibnamefont {O'toole}},\ }\href@noop {} {\bibfield  {journal} {\bibinfo
  {journal} {Trends in microbiology}\ }\textbf {\bibinfo {volume} {9}},\
  \bibinfo {pages} {34} (\bibinfo {year} {2001})}\BibitemShut {NoStop}%
\bibitem [{\citenamefont {Rodesney}\ \emph
  {et~al.}(2017{\natexlab{b}})\citenamefont {Rodesney}, \citenamefont {Roman},
  \citenamefont {Dhamani}, \citenamefont {Cooley}, \citenamefont {Touhami},\
  and\ \citenamefont {Gordon}}]{RN12195}%
  \BibitemOpen
  \bibfield  {author} {\bibinfo {author} {\bibfnamefont {C.~A.}\ \bibnamefont
  {Rodesney}}, \bibinfo {author} {\bibfnamefont {B.}~\bibnamefont {Roman}},
  \bibinfo {author} {\bibfnamefont {N.}~\bibnamefont {Dhamani}}, \bibinfo
  {author} {\bibfnamefont {B.~J.}\ \bibnamefont {Cooley}}, \bibinfo {author}
  {\bibfnamefont {A.}~\bibnamefont {Touhami}}, \ and\ \bibinfo {author}
  {\bibfnamefont {V.~D.}\ \bibnamefont {Gordon}},\ }\href {\doibase
  10.1073/pnas.1703255114} {\bibfield  {journal} {\bibinfo  {journal} {Proc
  Natl Acad Sci U S A}\ }\textbf {\bibinfo {volume} {114}},\ \bibinfo {pages}
  {5906} (\bibinfo {year} {2017}{\natexlab{b}})}\BibitemShut {NoStop}%
\bibitem [{\citenamefont {Witten}\ and\ \citenamefont
  {Pincus}(2010)}]{witten2010structured}%
  \BibitemOpen
  \bibfield  {author} {\bibinfo {author} {\bibfnamefont {T.~A.}\ \bibnamefont
  {Witten}}\ and\ \bibinfo {author} {\bibfnamefont {P.~A.}\ \bibnamefont
  {Pincus}},\ }\href@noop {} {\  (\bibinfo {year} {2010})}\BibitemShut
  {NoStop}%
\end{thebibliography}%
\newpage
\section*{Supplemental Information}

\section*{Other mechanical measurements}
We performed independent mechanical measurements \cite{RN13188,forgacs1998viscoelastic}, and found that the Young’s moduli were $14.8 \pm .8$ kPa and $3.1 \pm 1.3$ kPA (for Matrix+ and Matrix-, respectively). We found that the Matrix+ biofilms had an average surface tension of $0.089 \pm 0.001$ N/m, as compared to Matrix- biofilms’ average of $0.023 \pm 0.004$ N/m, and bending rigidities of $2.5 \pm 0.20 \times 10^{-9}$ Nm (Matrix+) vs. $3.4 \pm 0.45 \times 10^{-11}$ Nm (Matrix-).
\par
\section*{Tracer beads in Matrix+ biofilms}

\begin{figure}
\centering
    \includegraphics[width=\linewidth]{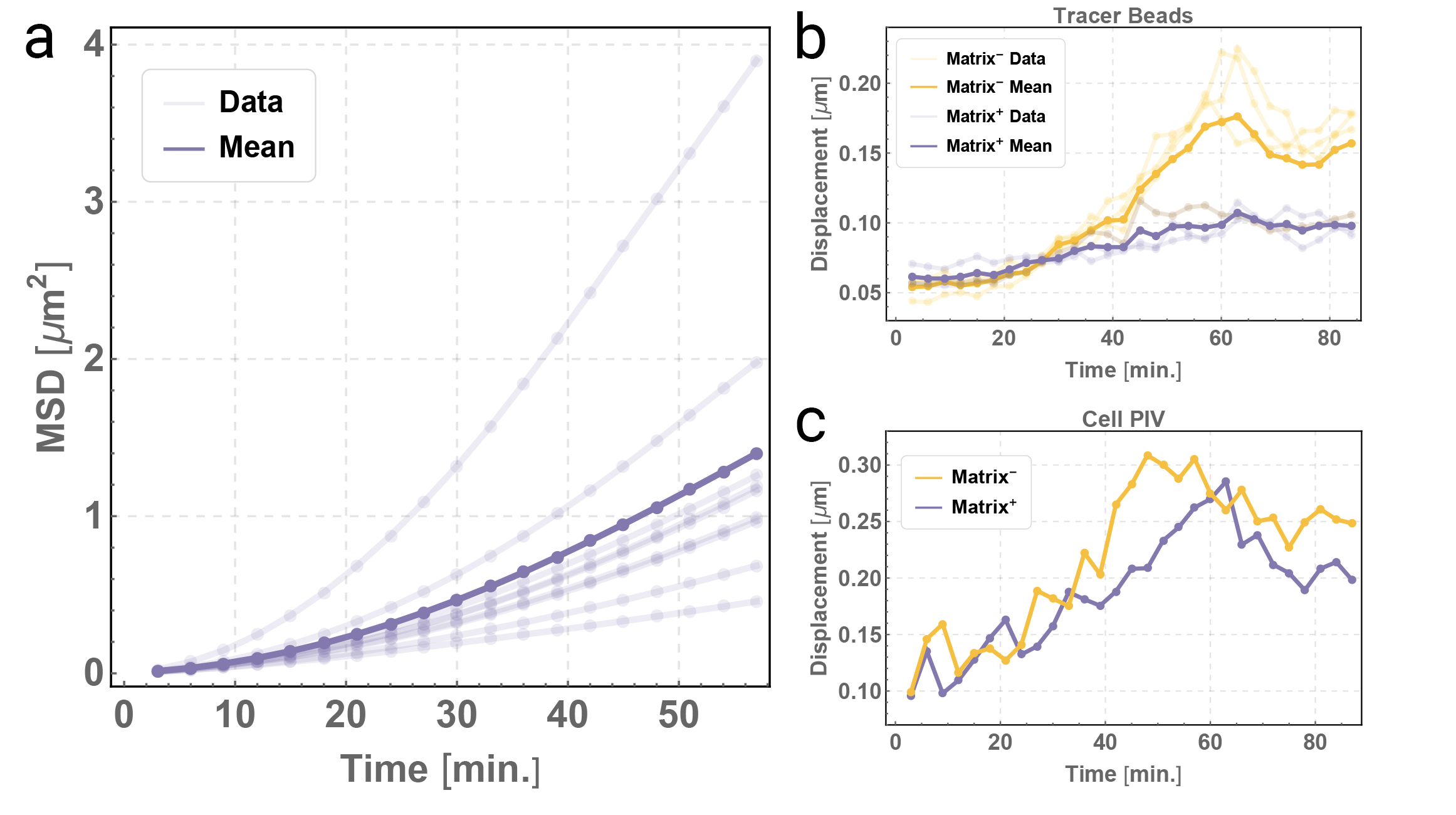}
    \caption{a) Measured lateral MSD for tracer beads inside a Matrix+ biofilm. Light lines represent data from  individual measurements; the dark line represents the average MSD. b) The displacement of the beads at early times during biofilm formation in Matrix- and Matrix+ biofilms shows that there is a strong divergence in bead motion after about 40 minutes of growth. c) However, using PIV to track individual cells in Matrix- and Matrix+ biofilms shows no such discrepancy, suggesting that the presence of an extracellular matrix prevents tracer beads from accurately tracing cellular motion.}
    \label{fig:mpMSD}
\end{figure}

In addition to Matrix- biofilms, we have extracted diffusion coefficients from tracer beads embedded in Matrix+ biofilms. We observed that  the MSD in Matrix+ biofilms also exhibits a caged-like plateau at short times, followed by diffusive-like linear regime at long lag times (figure \ref{fig:mpMSD}a). However, the variation in motion from sample-to-sample was higher in Matrix+ biofilms than Matrix- biofilms. We found that the diffusion coefficient was $0.16 \pm 0.11 \times 10^{-3} \mu m^2/s$, smaller than that in Matrix- biofilms ($\textit{p}=4 \times 10^{-4}$) .

To determine if the presence of extracellular matrix prevents tracer beads from accurately mimicking the motion of cells we need to compare the motion of cells and tracer beads. To measure the motion of individual cells, we analyzed the dynamics at early times during biofilm formation, i.e., these experiments started with 14\% of the agar surface covered with cells (and 1\% of the surface covered with beads), and continued until a dense bacterial monolayer formed. We measured the average displacement of the beads and the cells over time with time steps of 3 min. At very low densities, the beads behave similarly in Matrix+ and Matrix- biofilms (figure \ref{fig:mpMSD}b). However, after about 30 min, which corresponds to about 40\% of the surface covered with cells, the average displacements start to differ. Beads in Matrix- biofilms move faster, which is consistent with the difference in the diffusion coefficients extracted from tracer bead MSDs. Next, we analyzed the average displacement of cells using particle image velocimetry (PIV) during the same time span. We found that motion of cells in Matrix+ and Matrix- bioflims is not significantly during this time span (figure \ref{fig:mpMSD}c). This clearly contrasts with the behavior of beads shown in figure \ref{fig:mpMSD}b. In Matrix- biofilms, the displacement of beads and cells is similar at later times (beyond 40 min). We can conclude that the presence of extracellular matrix impacts the motion of tracer beads, as previously demonstrated \cite{birjiniuk2014single}. Thus, tracer beads do not accurately mimic the motion of cells in Matrix+ biofilms.

\section*{Langevin equation}
\subsection*{Effective temperature and related work}
As discussed in the main text, others have studied cell migratory behavior due to death and reproduction, and have used an effective temperature formulation to recover fluctuation-dissipation like behavior. In particular, we build on the work of Ranft, \textit{et al.}, \cite{RN12155} and Risler, \textit{et al.}, \cite{RN12359}. We use an effective temperature which agrees with theirs to leading order, but in this section we rewrite it in a different form for convenience. Ranft, \textit{et al.}, report an effective temperature that scales as: 
\begin{equation}
    k_B T_{\textrm{eff}} \propto \frac{\lambda_{\textrm{act}} \eta }{\rho},
    \label{Eq:ranft_teff}
\end{equation}
with the cellular activity rate $\lambda_{\textrm{act}}$, the effective viscosity $\eta$, and the cell number density $\rho$. Note that to our knowledge, previous work has focused on elastic cells that develop an \textit{effective} viscosity due to the fluidization effect of activity. Here, we'll be using, as $\eta$, the measured viscosity of biofilms. Viscosity is nothing more than a material's energy density multiplied by a timescale associated with structural rearrangements \cite{witten2010structured}. We can write $\eta = \frac{u}{V} \frac{1}{\gamma}$, for an energy $u$, a volume of material $V$, and a relaxation rate $\gamma$. Next we can write the inverse cellular number density as $\frac{1}{\rho} = \frac{V}{N}$ for $N$ cells. Combining these terms in equation (\ref{Eq:ranft_teff}), we have: 

\begin{align*}
    k_BT_{\textrm{eff}} &= \frac{\lambda_{\textrm{act}}}{\gamma} \cdot \frac{u}{N}
\end{align*}
\begin{equation}
    k_BT_{\textrm{eff}} = \langle E_{\textrm{const}} \rangle \cdot  \frac{\lambda_{\textrm{act}}}{\gamma} 
    \label{Eq:yanni_teff}
\end{equation}
where $\langle E_{\textrm{const}} \rangle$ describes the average energy per constituent, referred to as $U_0$ in the main text. For a passive material at equilibrium, the timescales associated with activity and structural rearrangement are equivalent (for example, they might both arise from molecular collisions), $\lambda_{act} = \gamma $, and the mean energy per constituent follows from equipartition $\langle E_{\textrm{const}} \rangle \propto k_BT$ (up to a constant depending on the number of degrees of freedom of the constituent). Therefore at equilibrium for a passive material $k_BT_{\textrm{eff}} = k_BT$, as one would hope. In our experiments $\langle E_{\textrm{const}} \rangle$ is associated with cellular motion, and so $T_{\textrm{eff}}$ measured in Kelvin is enormous compared to temperatures typically associated with atomic or molecular motion in daily experience.

\subsection*{Brief refresher on Brownian motion}
The traditional Langevin equation for Brownian motion can be used here, as described for instance in \cite{kubo1966fluctuation}. We start by writing Newton's second law for a particle in one dimension subject to viscous damping and a force which is random in time, $R(t)$. The trajectory of a particle will therefore be stochastic; we are restricted to investigating probability distributions and averages over many particles' trajectories, each with different realizations of the random force $R(t)$.
\begin{equation*}
    m\dot{v} = -m\gamma v + R(t),
\end{equation*}
where $R(t)$ is random but has the following known properties:
\begin{align*}
    \langle R(t) \rangle &= 0\\
    \langle R(t)R(t+\tau) \rangle &= \kappa \delta(\tau)
\end{align*}
where, in equilibrium for a passive fluid $\kappa = 2k_{\textrm{B}}Tm\gamma$. The angled braces $\langle ... \rangle$ denote averages over many realizations, and $\delta(...)$ is the Dirac delta function. We follow a typical derivation to find the diffusion constant, except that we leave our solution in terms of $T_{\textrm{eff}}$ instead of $T$. Again, for a passive fluid in equilibrium $T_{\textrm{eff}}$ simplifies to $T$. Dividing through by $m$ and letting $\xi(t) = \frac{R(t)}{m}$, yields
\begin{equation*}
    \dot{v} = -\gamma v + \xi(t),
\end{equation*}
which can be solved using standard techniques as 
\begin{align*}
    v(t) &= v_0e^{-\gamma t}+e^{\gamma t} \int_0^t{dt' e^{\gamma t'}\xi(t')} \\
\end{align*}
Without loss of generality (at least when finding diffusion constants), we can set $v(0) = 0$ and $x(0) = 0$ then
\begin{align*}
    v(t) &=e^{\gamma t} \int_0^t{dt' e^{\gamma t'}\xi{t')}} \\
    x(t) &= \int_0^t{  e^{\gamma t'} \int_0^{t'}{dt'' e^{\gamma t''}\xi(t'')}  }
\end{align*}

Squaring, taking the average over many solutions (i.e. for many realizations of the stochastic function $\xi$), and using the relation $\langle \xi(t) \xi(t+\tau) \rangle = \frac{2\gamma k_B T_{\textrm{eff}}}{m} \delta(\tau)$ yields (after much rearranging):

\begin{align}
    \langle v^2 \rangle &= \frac{k_{\textrm{B}} T_{\textrm{eff}}}{m}(1-e^{-2\gamma t}) 
    \label{Eq:eq_all_time1}\\
    \langle x^2 \rangle &= \frac{ k_{\textrm{B}} T_{\textrm{eff}}}{2m\gamma^2} \left( 2\gamma t - 3 + 4e^{-\gamma t} - 2e^{-2\gamma t} \right)
    \label{Eq:eq_all_time2}
\end{align}

In the long-time limit this becomes
\begin{align}
    \langle v^2 \rangle &= \frac{k_{\textrm{B}} T_{\textrm{eff}}}{m} & \text{Equipartition}
    \label{Eq:equipartition}\\
    \langle x^2 \rangle &= \frac{ k_{\textrm{B}} T_{\textrm{eff}}}{m\gamma}t &\text{Diffusion}
    \label{Eq:diffusion}
\end{align}

\subsection*{Conventional diffusion and viscosity independent diffusion}
As discussed above, eq (\ref{Eq:yanni_teff}) simplifies to $k_{\textrm{B}}T$ for a passive fluid at equilibrium; replacing $k_{\textrm{B}}T_{\textrm{eff}}$ with $k_{\textrm{B}}T$ in eq (\ref{Eq:diffusion}) yields the expected result for passive, equilibrium systems. This is because we've set the timescale for damping equal to the timescale associated with thermal kicks --- $\gamma$ appears in the noise strength as well as in the damping term. Also, the energy of each constituent is again set by the strength of thermal kicks, thus ensuring equipartition holds.

On the other hand, when the timescales governing the source of constituent motion and damping of constituent motion are separated, Einstein's classic results no longer necessarily hold. In our case, when considering active, reproducing but immotile, constituents embedded in a viscoelastic medium, it is most natural to set the inverse damping timescale $\gamma$ to the viscoelastic relaxation rate  $\frac{E}{\eta}$. The energy scale, however, is associated with cellular motions due to \textit{active} kicks (i.e. step stresses from reproduction and death), and so we can write $\langle E_{\textrm{const}} \rangle =m\left(\frac{d \sigma_0}{\eta}\right)^2$, with cellular diameter $d$. The reproduction/death rate is identified with the activity rate $\lambda_{\textrm{const}}$. We then have

\begin{align*}
    \langle x^2 \rangle = \frac{k_{\textrm{B}}T_{\textrm{eff}}}{m \gamma}t \\
    \langle x^2 \rangle = \frac{\langle E_{\textrm{const}} \rangle \lambda_{\textrm{act}}}{m \gamma^2}t \\
    \langle x^2 \rangle = \frac{m\left(\frac{d \sigma_0}{\eta}\right)^2 \lambda_{\textrm{act}}}{m \left(\frac{E}{\eta}\right)^2}t \\
    \langle x^2 \rangle = \left( \frac{d \sigma_0}{E}\right)^2 \lambda_{\textrm{act}}t \\
    \langle x^2 \rangle = \ell^2 \lambda_{\textrm{act}}t \\
\end{align*}

Here $\ell = d \frac{\sigma_0}{E}$ is the change in length that the spring in the Voigt-Kelvin element of length $d$ would experience instantaneously if subjected to a force $\sigma_0$. In other words, we can think of it as the ``step-size" in a discrete one-dimensional random walk, where steps are taken at a rate $\lambda_{\textrm{act}}$. The above is the long-time result. The equation we use to compare with simulations at all times applies this same process to eq (\ref{Eq:eq_all_time2}), producing:

\begin{equation}
    \langle x(t)^2 \rangle = \frac{\lambda_{\textrm{act}}\ell^2}{2\gamma}\left( 2\gamma t - 3 +4e^{-\gamma t} - 2e^{-2\gamma t} \right)
    \label{Eq:prediction}
\end{equation}

\section*{Simulations}
Biofilm simulations in which the reproduction and death rate is $\lambda_{\textrm{act}}$ and the viscoelastic relaxation rate is $\gamma = \frac{E}{\eta}$ agree well with the predicted mean squared displacement from eq \ref{Eq:prediction}.

As described in the main text, we performed event-driven, individual-based simulations of mutual killer cells in 1D. To capture the viscoelastic character of the biofilm, we model it as a chain of cells coupled by Voigt-Kelvin elements, with spring stiffness $E$ and dashpot damping $\eta$. Reproduction and death are assumed to be Poisson processes with an activity rate $\lambda_{\textrm{act}}$; the time-step between events is chosen from an exponential distribution $dt \sim e^{-\lambda_{\textrm{act}}t}$. Each event corresponds, with equal probability, to the step-strain resultant from reproduction or death of a cell at a random position in the biofilm, and as such imposes a step stress $\sigma_0$ felt instantaneously throughout the biofilm. Finally, after each event the velocities and positions of all the cells are updated according to the current state of stress in the biofilm and the constitutive equations $\sigma(t) = E\epsilon(t) + \eta \frac{d}{dt}{\epsilon}(t)$, using backward-Euler integration. This leads to the discretization scheme:
\begin{equation*}
\epsilon_t = \frac{1}{1 + dt\frac{E}{\eta}} \cdot (\epsilon_{t-1} + \frac{dt}{\eta}\sigma_t))
\end{equation*} In fact, the simulation code is very short so it is included inline here.

\begin{python}

### imports
import numpy as np
from matplotlib import pyplot as plt

### define constants (example values)
E = 1; eta = 1
T = 20000 #total number of time steps
lambda_act = 1e-2 #rate of birth/death events
N = 10000 #length of biofilm
sigma_0 = 10

### positions of events
j = np.random.randint(0,N,size=T)

###nature of events 
#(birth = 1, death = -1)
s = np.random.choice([-1,1], size=T)

###timing of events
dt = np.random.exponential(scale = 1/lambda_act, size = T)
t = np.cumsum(dt)

sigma = np.zeros(shape=(T,N))
epsilon = np.zeros(shape=(T,N))

### run the simulation
for k in range(T):
	sigma[k,:j[k]] += -sigma_0*s[k]
	sigma[k,j[k]:] += sigma_0*s[k]
	epsilon[k,:] = (1 / (1 + dt[k]*E/eta)) * (epsilon[k-1,:] + (dt[k]/eta)*sigma[k,:])
	
### plot results
plt.plot(t,np.average(epsilon**2,axis=1))
\end{python}

\section*{Physical Explanation of Langevin Modification}

The Langevin equation results eq (\ref{Eq:eq_all_time2}) with $k_{\textrm{B}}T_{\textrm{eff}}$ given by eq (\ref{Eq:yanni_teff}) agree with simulations at all times, but the connection between the simulations and the effective temperature approach remains perhaps un-intuitive. In this section we attempt to give a more mechanical picture of how the simulations relate to the Langevin equation approach.

\begin{figure}
\centering
    \includegraphics[width = .8 \linewidth]{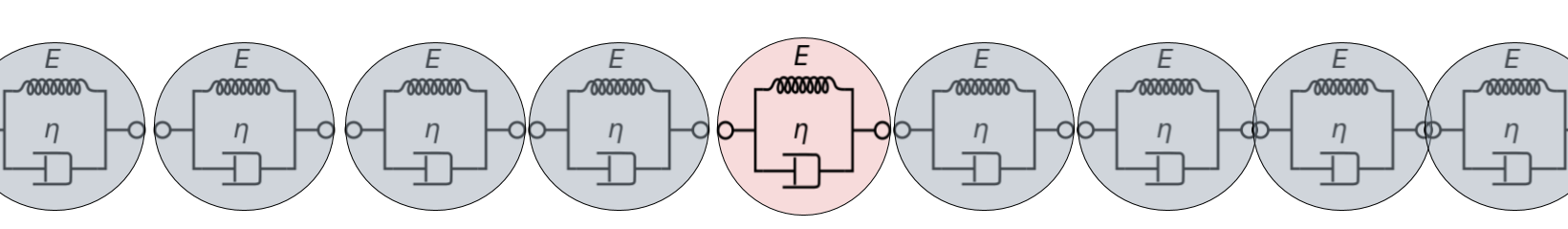}
    \caption{1D chain of Voigt-Kelvin elements}
    \label{fig:cubb}
\end{figure}

Our model of a biofilm is shown in figure \ref{fig:cubb}. It is a 1D chain of incompressible points separated by Voigt-Kelvin cells. When a new cell is created its mother shoves everybody aside with all her force, $f_{max}$. This force will get distributed through the chain as a compressive stress $\sigma_0$. When a cell lyses, it will leave behind a tensile stress in its wake. The magnitude of this stress we'll also set, for convenience, to $\sigma_0$. We consider only biofilms in the homeostatic limit so the rate of lysis and division balance. 

\begin{figure}
\centering
    \includegraphics[width=.85\linewidth]{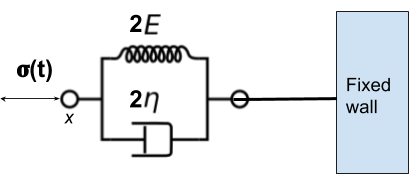}
    \caption{Equivalent view to tracking the pink cell}
    \label{fig:Darheel}
\end{figure}

Consider the motion of a cell somewhere near the center of a vast one-dimensional biofilm (the pink cell in figure 7). This cell is pushed and pulled, back-and-forth, in response to stresses as cells to its left and right divide and lyse. But how many stresses occur over a time $t$? If we assume that division and lysis events are Poisson distributed throughout the biofilm then we can write:
\begin{equation*}
\sigma_{\textrm{total}} = \sigma_0 \left(N^{\textrm{right}} - N^{\textrm{left}} \right)
\end{equation*}
where $N^{\textrm{right}}$ and $N^{\textrm{left}}$ are distributed as:
\begin{equation*}
    p(N) = e^{-\lambda_{\textrm{act}} t} \frac{(\lambda_{\textrm{act}} t)^N}{N!}
\end{equation*} with $\lambda_{\textrm{act}}$ the rate of division and lysis in the biofilm. The probability that the cell experiences $\sigma(t)$ at time $t$ will then be a Skellam distribution (which arises from the difference of two Poisson processes).
\begin{equation*}
    p(\frac{\sigma}{\sigma_0}) = e^{-2\lambda_{\textrm{act}}t}I_{|N|}(2\lambda_{\textrm{act}}t)
\end{equation*}
where $I_{|N|}(2\lambda_{\textrm{act}}t)$ is the modified Bessel function of the first kind. In the limit of large $\lambda_{\textrm{act}}$ the distribution converges to a Gaussian with mean $0$ and variance $\lambda_{\textrm{act}} t$. This is unsurprising as it's clear that we can think of the current value of the total stress as undergoing a random walk in one dimension. In other words, we flip a coin at a rate $\lambda_{\textrm{act}}$ and each time take a step $\pm \sigma_0$ based on whether a birth (lysis) event happened to our left (right). To get a particular trajectory we can write 
\begin{align*}
\dot{\sigma}(t) &= \sigma_0 \psi(t) \\
\sigma(t) &= \sigma_0 \int_0^{t}{\psi(t^{\prime}) dt^{\prime}}
\end{align*}
with 
\begin{align*}
\langle \psi(t) \rangle &= 0 \\
\langle \psi(t) \psi(t^{\prime}) \rangle &= \lambda_{\textrm{act}} \delta(t - t^{\prime}) \\
\end{align*}
which will give us the same distribution for $\sigma(t)$ in the end. This gives us a clearer physical picture of the cellular diffusion as well. Tracking a single cell's motion over time is equivalent to tracking the motion of point x in figure \ref{fig:Darheel}. This is nothing more than one much larger Voigt-Kelvin cell subject to a stress that randomly walks in time.

The equation of motion for x is then easy to write down:

\begin{align*}
\sigma(t) &= E\epsilon(t) + \eta \dot{\epsilon}(t) \\
\sigma_0 \int_0^{t}{\psi(t^{\prime}) dt^{\prime}} &= E\epsilon(t) + \eta \dot{\epsilon}(t) \\
E \epsilon_0 \int_0^{t}{\psi(t^{\prime}) dt^{\prime}} &= E\epsilon(t) + \eta \dot{\epsilon}(t) \\
\lambda_\eta \epsilon_0 \int_0^{t}{\psi(t^{\prime}) dt^{\prime}} &= \lambda_\eta \epsilon(t) + \dot{\epsilon}(t) \\
\lambda_\eta \ell \int_0^{t}{\psi(t^{\prime}) dt^{\prime}} &= \lambda_\eta x(t) + \dot{x}(t) \\
\lambda_\eta \ell \psi(t) &= \lambda_\eta v(t) + \dot{v}(t) \\
\end{align*}

here, $\sigma_0$ is written as $E\epsilon_0$, where $\epsilon_0$ is how much the spring \textit{would} be strained if it were disconnected from the Voigt-Kelvin cell and subjected on its own to an instantaneous stress $\sigma_0$. Then we divided through by $\eta$ to present everything in terms of a viscoelastic relaxation rate $\lambda_\eta$. Next, we multiplied by a cell's relaxed length $x_0$, which turns $\epsilon$ into $x$ and $\dot{\epsilon}$ into $\dot{x}$. Note that $\epsilon_0 \cdot x_0 = \ell$. Finally, we take a time derivative to get an equation similar to the familiar Langevin equation for Brownian motion. The final line would yield the usual fluctuation-dissipation relation, except that there's a term multiplying the white noise. The root mean squared velocity and position can be solved for; this is the same exercise as above, and yields:

\begin{equation*}
    \langle x(t)^2 \rangle = \frac{\lambda_{\textrm{act}}\ell^2}{2\lambda_\eta}\left( 2\lambda_\eta t - 3 +4e^{-\lambda_\eta t} - 2e^{-2\lambda_\eta t} \right)
\end{equation*}

\begin{align*}
\langle x(t)^2 \rangle &= \lambda_{\textrm{act}} \ell^2 t &
t \rightarrow \text{Large}
\end{align*}

\end{document}